\documentclass[%
aip,
 apl,
 amsmath,amssymb,
 reprint,%
]{revtex4-1}

\usepackage{graphicx}
\usepackage{dcolumn}
\usepackage{bm}
\usepackage[english]{babel}
\usepackage[utf8]{inputenc}
\usepackage[T1]{fontenc}
\usepackage{xcolor}
\usepackage{ulem}
\usepackage{comment}
\usepackage{mathtools}

\begin{document}

\title{Probing anisotropy in epitaxial Fe/Pt bilayers by spin-orbit torque ferromagnetic resonance}

\author{Mohammad Tomal Hossain}
\affiliation{Department of Physics and Astronomy, University of Delaware, Newark, Delaware 19716, USA}%
 \author{Sergi Lendinez}
\affiliation{Department of Physics and Astronomy, University of Delaware, Newark, Delaware 19716, USA}%
\altaffiliation[Currently at ]{Center for Advanced Microstructures and Devices, Louisiana State University, Baton Rouge, LA 70806, USA.}

\author{Laura Scheuer}
\affiliation{Fachbereich Physik and Landesforschungszentrum OPTIMAS, Technische Universit\"at Kaiserslautern, 67663 Kaiserslautern, Germany}

\author{Evangelos Papaioannou}
\affiliation{Institut f\"ur Physik, Martin-Luther-Universit\"at Halle-Wittenberg, Halle, Germany}%
\author{M. Benjamin Jungfleisch}
\email[]{mbj@udel.edu}
\affiliation{Department of Physics and Astronomy, University of Delaware, Newark, Delaware 19716, USA}%

\date{\today}

\begin{abstract}

We report the generation and detection of spin-orbit torque ferromagnetic resonance (STFMR) in micropatterned epitaxial Fe/Pt bilayers grown by molecular beam epitaxy. The magnetic field dependent measurements at an in-plane magnetic field angle of 45$^\circ$ with respect to the microwave-current direction reveal the presence of two distinct voltage peaks indicative of a strong magnetic anisotropy. We show that STFMR can be employed to probe the underlying magnetic properties including the anisotropies in the Fe layer. We compare our STFMR results with broadband ferromagnetic resonance spectroscopy 
 of the unpatterned bilayer thin films. The experimental STFMR measurements are interpreted using an analytical formalism and further confirmed using micromagnetic modeling, which shed light on the field-dependent magnetization alignment in the microstructures responsible for the STFMR rectification. Our results demonstrate a simple and efficient method for determining magnetic anisotropies in microstructures by means of $rf$ spectroscopy.

\end{abstract}

\maketitle


Exploring spin-orbit torques (SOTs) in novel material systems is a prosperous field in spintronics that has attracted enormous attention in the past decade. Studies on SOTs  enable both the realization of highly energy efficient storage applications and an improvement of our understanding of fundamental spin physics at interfaces. In this regard, spin-torque or spin-orbit torque ferromagnetic resonance (STFMR) is a prominent choice for studying spin-orbit toques in multilayers \cite{Liu_PRL2011,Harder_PRB2011}. 
It has been demonstrated that this mechanism can be observed in a large range of material systems including metallic \cite{Liu_PRL2011} and insulating ferromagnetic bilayers \cite{Jungfleisch_PRL2016,Sklenar_PRB2015},  antiferromagnets \cite{Zhang_PRB2015}, heterostructures made of topological insulators \cite{Mellnik2014} and transition metal dichalcogenides \cite{Zhang_APLMaterials}, etc. A particular emphasis of these efforts has been on determining the spin-to-charge conversion efficiency and the spin-Hall angle. Moreover, most of these works relied on a macro-spin approach without considering any anisotropies \cite{Chiba_PRAppl,Liu_PRL2011}. 

An overlooked aspect of STFMR has been the dynamic response driven by oscillatory SOT. Other magnon spintronic effects such as spin pumping, spin-Hall effect and auto-oscillation studies have extensively been used to determine what role the underlying magnetization dynamics plays. For instance, Sandweg et al. reported an enhanced contribution of surface spin-wave modes to the spin pumping signal \cite{Sandweg_APL}. Meanwhile, Papaioannou et al. observed strong magnetic anisotropies giving rise to two distinct inverse spin-Hall effect voltage peaks driven by spin pumping in Fe/Pt \cite{Papaioannou_APL2013}. Corresponding reports on STFMR driven dynamics are scarce \cite{Fulara,Zhou_2019} and successful demonstration of probing magnetic anisotopy directions by STFMR remained elusive until now.

Here, we report the generation and detection of STFMR in micropatterned epitaxial Fe/Pt bilayers grown by molecular beam epitaxy. We compare our STFMR results with standard ferromagnetic resonance spectroscopy and show that in-plane magnetic field dependent STFMR can be employed to probe all underlying magnetic properties including the anisotropies in the epitaxial Fe layer. The experimental results are interpreted using an analytical formalism and further confirmed using micromagnetic modeling. Our results demonstrate a simple and efficient method for determining magnetic anisotropies in microstructures by means of $rf$ spectroscopy.




\begin{figure}[t]
    \centering
    \includegraphics[width=0.90\columnwidth]{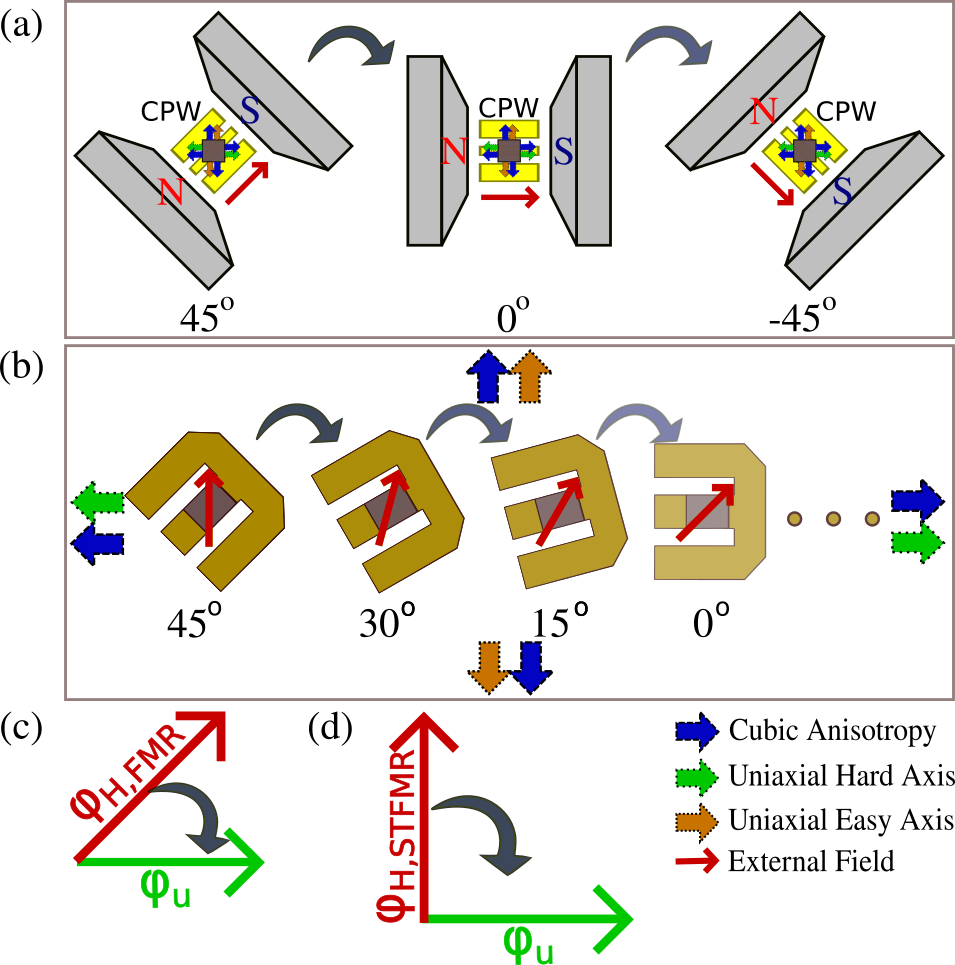}
    \caption{Illustration of experimental setup and measurement configuration. (a) Schematic illustration of flip-chip FMR measurement configuration of the unpatterned Fe/Pt films using a coplanar waveguide (CPW). The films were placed on the top of a CPW parallel to the external magnetic field. Measurements for three different in-plane field directions ($\theta_\mathrm{FMR}=45^\circ, 0^\circ, -45^\circ$) were collected. (b) Illustration of the STFMR measurement configuration with the patterned devices. The Fe/Pt microstructures have lateral dimensions of 80 $\mu$m x 130 $\mu$m. 
    A series of samples with different orientations in $15^\circ$-steps with respect to crystallographic directions ($\theta_\mathrm{STFMR}=45^\circ, 30^\circ, 15^\circ, 0^\circ ...$) are patterned. The solid straight arrows show the magnetic anisotropy and the in-plane magnetic field directions. Schematic of the coordinate systems used for (c) FMR and (d) STFMR. The red arrows indicate the orientations of the corresponding in-plane fields which rotates clockwise by $45^\circ$ (FMR) or $15^\circ$ (STFMR) (please note that for the FMR measurements $\varphi_\mathrm{H,FMR}=\theta_\mathrm{FMR}$ and for the STFMR measurements $\varphi_\mathrm{H,STFMR}=45^\circ + \theta_\mathrm{STFMR}$). $\varphi_\mathrm{u}$ is the hard uniaxial anisotropy direction (easy axis direction perpendicular to $\varphi_\mathrm{u}$) which remains fixed at $0^\circ$ in our coordinate setup. All angles are measured with respect to the horizontal direction.}
    \label{fig:fig1}
\end{figure}
{The bilayers were grown epitaxially on MgO ($100$) substrates by electron-beam evaporation under a base pressure of $5\times 10^{-9}$ mbar. The substrate’s surface was initially cleaned in organic solvents, followed by annealing at 650$^{\circ}$C for 1 hour. The bilayers were then deposited with a rate of 0.01~nm/s at a substrate temperatures 300$^{\circ}$C, followed by 1 hour total annealing time at the growth temperature, resulting in a Fe ($10$~nm)/ Pt ($5$~nm) bilayer configuration.} 

{The epitaxial films were then etched to produce the STFMR devices: first, a negative-tone photoresist 
was used to cover 80 $\mu$m $\times$ 130~$\mu$m rectangular sections of the film. 
The rectangular sections are rotated 15$^\circ$ to cover relative angles between $\theta_\textrm{STFMR}=45^\circ$ and $\theta_\textrm{STFMR}=-120^\circ$. The surrounding film was then etched using Ar ion milling. 
The shorted coplanar waveguides were patterned on the top of the rectangular epitaxial film sections using a direct laser writer and a positive-tone photoresist. 10~nm of Ti and 100~nm of Au were e-beam evaporated and lifted off to finalize the waveguide contacts on the devices; see Fig.~\ref{fig:fig1} for a schematic}

We introduce two angular coordinate sets for our setup: $\theta$, the sample/device orientation angle, and $\varphi$, the field angle; both defined with respect to the anisotropy axes [see Fig.~\ref{fig:fig1}]. For both FMR and STFMR measurements, the coordinate system is defined with respect to the hard axis of the uniaxial anisotropy of the sample. In our Fe/Pt cubic system, an in-plane fourfold magnetic anisotropy is expected due to the cubic lattice symmetry of Fe, together with an additional uniaxial magnetic anisotropy, which is superimposed on top of the fourfold anisotropy \cite{KARFARIDIS,Zhan_2009}.
The FMR measurements were performed prior to the microstructuring in STFMR devices using the identical samples. For the FMR measurements, the entire substrate is rotated anti-clockwise in steps of $45^{\circ}$ for successive measurements. In our coordinate setup (where hard axis of the uniaxial anisotropy is fixed along x-axis), the magnetic field is effectively rotated clockwise by $45^{\circ}$. We set the sample angle $\theta_\mathrm{FMR}=0^{\circ}$ when the external magnetic field is parallel to the hard axis [see Fig. \ref{fig:fig1}(a)]. 
The external field angle $\varphi_\mathrm{H,FMR}$ is rotated clockwise with respect to the hard axis of the uniaxial anisotropy [see Fig. \ref{fig:fig1}(c)], so that $\varphi_\mathrm{H,FMR}=\theta_\mathrm{FMR}$. 
The bilayers were then patterned into microstrips for STFMR experiments at orientations ($\theta_\mathrm{STFMR}$) clockwise with respect to the hard axis 
in steps of $15^{\circ}$ as illustrated in Fig. \ref{fig:fig1}(b). We set $\theta_\mathrm{STFMR}=0^{\circ}$ when the pattern is parallel to the hard axis of the uniaxial anisotropy field and assign the device angles $\theta_\mathrm{STFMR}$ with respect to this position [see Fig.~\ref{fig:fig1}(b)]. The external magnetic field [shown as a red arrow in Fig.~\ref{fig:fig1}(b)] for all measurements is applied at $45^{\circ}$ with respect to the device edge so that a maximum STFMR signal strength can be achieved in saturation\cite{HARDER_Physrep2016,Harder_PRB2011,Liu_PRL2011}. Hence, the external field angle is given by $\varphi_\mathrm{H,STFMR}=45^{\circ}+\theta_\mathrm{STFMR}$. The fourfold cubic anisotropy direction is always along the substrate edges [blue arrows in Figs.~\ref{fig:fig1}(a,b)]. In this coordinate setup it is easy to realize that the FMR device oriented at $\theta_\mathrm{FMR}$ is equivalent to the STFMR configuration oriented at $\theta_\mathrm{STFMR}=\theta_\mathrm{FMR}-45^{\circ}$. All measurements were performed at room temperature. 

For the FMR measurements [Fig.~\ref{fig:fig1}(a)], the magnetization dynamics is excited by the microwave magnetic field of a coplanar wave guide (CPW) in transmission. The transmitted {S$_{12}$ parameter is measured using a vector network analyzer}. 
The external magnetic field is applied in-plane and swept between $-2000$ Oe and $2000$ Oe. Upon achieving the FMR resonance condition [Eq.~(\ref{eq.  -b}) below], the microwave magnetic field induces a precession of the magnetic moments in the Fe layer. This leads to a resonant absorption of the microwave signal and thus results in a characteristic symmetric Lorentzian absorption when approaching resonance \cite{Kalarickal_2006}. Compared to the bare Fe film, the Gilbert damping of the Fe/Pt bilayer is enhanced due to spin pumping from the Fe layer into the adjacent Pt layer \cite{Tserkovnyak_PRL2002,Papaioannou_APL2013}.


For the STFMR measurements a bias tee is used to simultaneously apply a microwave current and to measure the rectified $dc$ voltage using a lock-in amplifier. A microwave frequency signal of $22$ dBm power is supplied by an Agilent E8257D signal generator. 
The external magnetic field $H_\mathrm{ext}$ applied in the sample plane and swept between $-2000$ Oe and $2000$ Oe for each device.

\begin{figure*}[t]
    \centering
    \includegraphics[width=0.8\paperwidth]{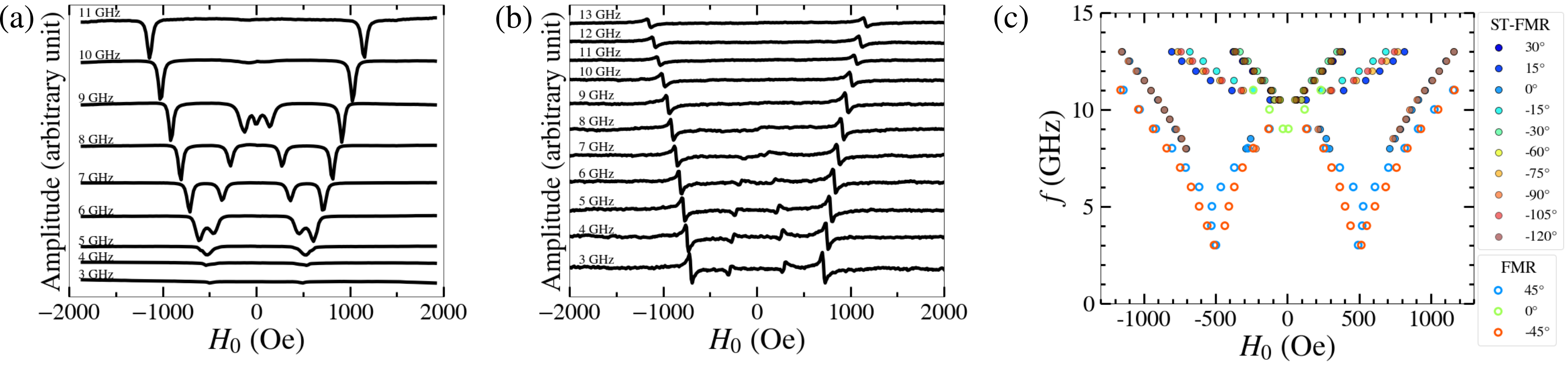}
    \caption{(a) Absorption spectrum obtained by broadband FMR at $\theta_\mathrm{FMR}=45^\circ$. (b) Corresponding results of the STFMR measurements of the microstructure at an orientation of $\theta_\mathrm{STFMR}=0^\circ$ (note that due to the chosen coordinate system $\theta_\mathrm{STFMR}=\theta_\mathrm{FMR}-45^{\circ}$). (c) Resonant frequency vs. magnetic field plot extracted from the FMR and STFMR results shown in (a) and (b). Open circles represent results obtained from FMR,  filled circles represent results obtained from STFMR measurements.}
    \label{fig:fig2}
\end{figure*}

The micromagnetic simulations are carried out using the graphics processor unit (GPU)-accelerated program Mumax3\cite{Vansteenkiste2014}. The device is modelled into $1024 \times 1024 \times 1$ cells with an individual cell size of $3.0$~nm~$\times~ 3.0$~nm~$ \times~5.0$~nm with periodic boundary conditions in two dimensions. The material parameters employed in simulations were obtained from the experimental data [fits to Eqs. (\ref{eq.  -a}) and (\ref{eq.  -b})]. The Gilbert damping constant\cite{Guillemard_APL2018} $\alpha = 0.0081$ and the exchange stiffness constant\cite{Goryunov_Phys.Rev.B1995} $A = 2\times 10^{-6}$ erg/cm were used as simulation parameters. An $ac$ {$\mathrm{sinc}$ pulse} driving magnetic field of amplitude 5 Oe and $50$ GHz cut-off frequency is applied at an angle $90^\circ+\theta_\mathrm{sim}$ with respect to the hard axis of the uniaxial anisotropy. A sweeping external field varied between 0 Oe and 2000 Oe is applied in the plane at $-45^\circ$ ($45^\circ$ in clockwise orientation) angle from the $ac$ driving field, hence $\varphi_\mathrm{H,sim}=45^\circ+\theta_\mathrm{sim}$. Thus, $\theta_\mathrm{sim}$ is equivalent to $\theta_\mathrm{STFMR}$. The magnetization is relaxed and simulated for a total duration of 3 ns without an $ac$ driving field to find the ground state configuration of the magnetization. The simulation is then run for another 4 ns for the dynamics simulation. The resonance frequency is then found from the Fourier transformation of the time dependence of the magnetization.




\begin{figure}[b]
    \centering
    \includegraphics[width=0.82\columnwidth]{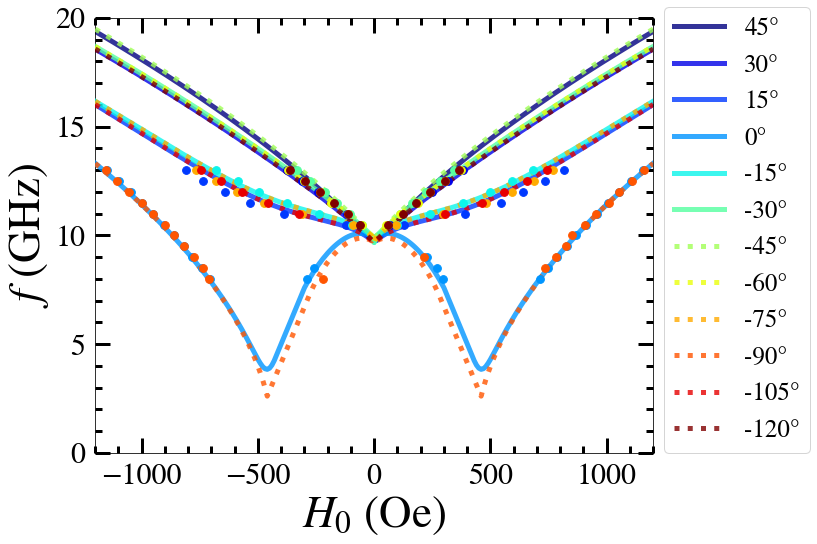}
    \caption{Fit to analytical model for an in-plane geometry, Eq.~(\ref{eq.  -b}). The solid and dotted lines show the fitted curves for different devices and the solid circles are the experimental results for STFMR measurement. The global minimum of the least square residue is numerically found and used as fitting method to fit all experimental STFMR data points for different device angles $\theta_\mathrm{STFMR}$ simultaneously. The obtained parameters are summarized in Tab.~\ref{fit table}.}
    \label{fig:fig3}
\end{figure}

Figure \ref{fig:fig2}(a) shows the FMR spectra ($S_{12}$ parameter) at $\theta_\mathrm{FMR}=45^{\circ}$ 
for different frequencies. The external field is swept between $-2000$~Oe to $2000$~ Oe at each frequency. Upon achieving the resonance condition, a minimum transmission ($S_{12}$) is observed, corresponding to a maximum absorption. The spectrum is symmetric with respect to zero field. 
At low frequencies (below 5~GHz), only one mode is detectable. As the frequency increases, a second peak emerges, see Fig.~\ref{fig:fig2}(a). This second peak decreases in field as the frequency is increased [Figs.~\ref{fig:fig2}(a) and (b)]. At even higher frequency ($f>$ 9.5 GHz), this second mode disappears and the remaining mode shows a typical Kittel-like increase of the resonance field with frequency. 


Figure \ref{fig:fig2}(b) shows the frequency-dependent STFMR results for a device aligned at $\theta_\mathrm{STFMR}=0^{\circ}$ [see Fig.~\ref{fig:fig1}(b)]. 
An oscillatory pure spin current is generated in the Pt layer by means of the spin-Hall effect \cite{Hirsch_PRL1999,Dyakonov1971}. 
Upon injection of this pure spin current in the Fe layer, it interacts with the magnetic moments in the ferromagnet by exerting a SOT, which results in the onset of a precession of the moments around an equilibrium axis. In addition, the Oersted field created by the alternating charge current contributes to the precession. Due to magnetoresistance effects, this results in an oscillatory resistance change that mixes with the microwave current leading to a rectified $dc$ signal.
This rectified $dc$ voltage is then detected by a lock-in amplifier through a bias tee\cite{Liu_PRL2011}. As is shown in Fig.~\ref{fig:fig2}(b), we observe two peaks for each frequency for positive and negative biasing field. The peaks separate  from each other for increasing frequency, in agreement with the FMR results shown in Fig.~\ref{fig:fig2}(a). The observed FMR and STFMR lineshapes are different as fundamentally different physical mechanisms lead to the signal observed on resonance \cite{Harder_PRB2011}.

To compare the FMR and STFMR measurements [Fig.~\ref{fig:fig2}(a) and Fig.~\ref{fig:fig2}(b)], we plot the frequency $(f)$/ field $(H_\mathrm{0})$ relationships obtained from both techniques at different angles, see Fig.~\ref{fig:fig2}(c). 
The resonance conditions were determined from fits to Lorentzian lines in the FMR and STFMR spectra [Fig.~\ref{fig:fig2}(a,b)]. For any given frequency/ field combination the results of a STFMR measurement for a device at $\theta_\mathrm{STFMR}$ qualitatively matches  that of the FMR measurement at $\theta_\mathrm{STFMR}=\theta_\mathrm{FMR}-45^{\circ}$ as expected from the definition of our coordinate system [see Fig. \ref{fig:fig1}]. 
This result shows that STFMR can be used to determine magnetic anisotropies in microstructures. 
For example, two distinct resonances are found when the external field is parallel to the cubic hard axis (i.e., $\theta_\mathrm{STFMR}=0^{\circ}$ and $-90^{\circ}$ and $\theta_\mathrm{FMR}=45^{\circ}$ and $-45^{\circ}$). The physical reason behind this can be explained through resonance conditions as detailed in the following.
    
\begin{figure}[t]
    \centering
    \includegraphics[width=0.82\columnwidth]{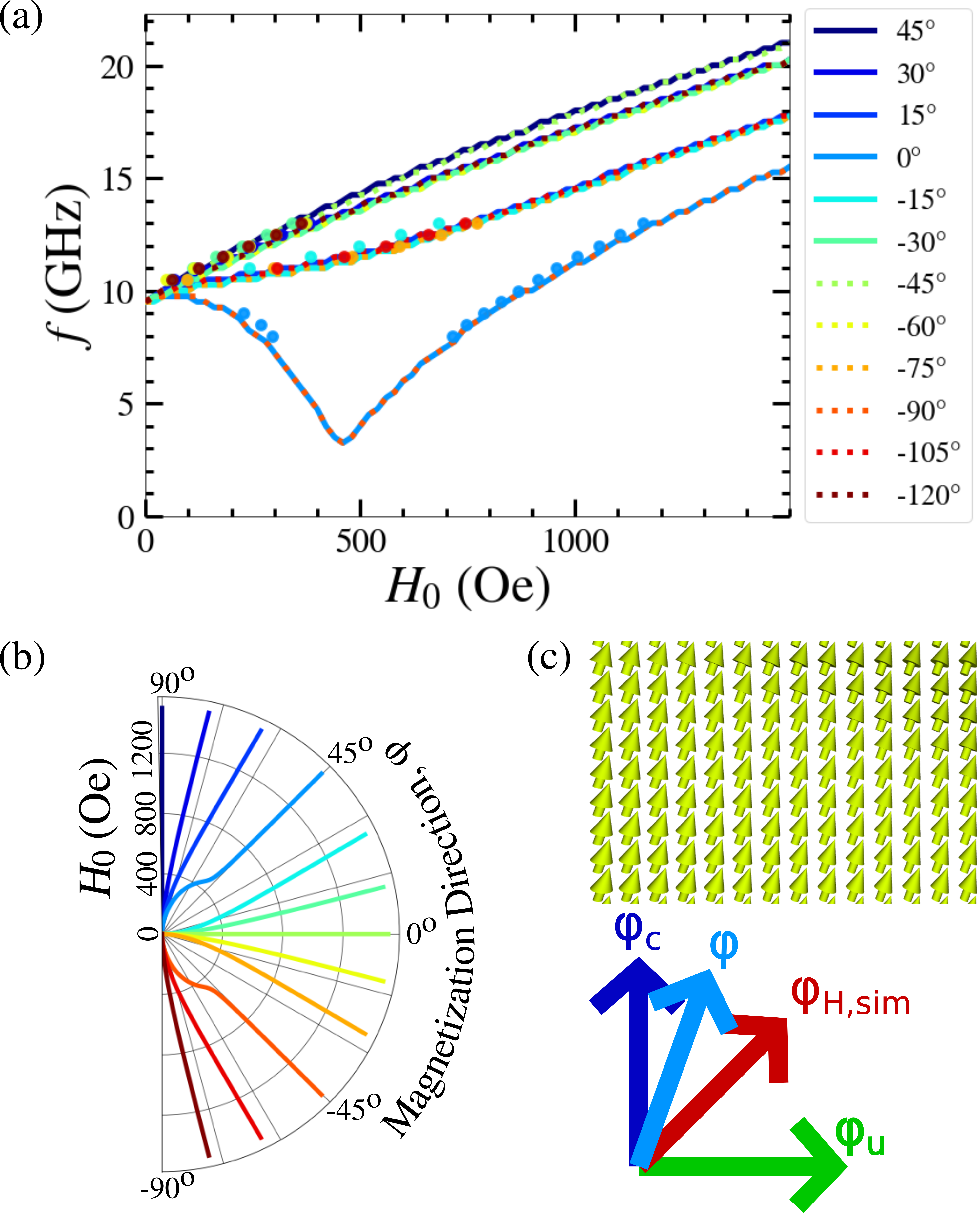}
    \caption{(a) Frequency vs. resonating field for different $\theta_\mathrm{sim}$ from micromagnetic modeling using Mumax3. The magnetic parameters were extracted from experimental data shown in Fig. \ref{fig:fig3}. (b) Polar plot of the magnetization direction $\varphi$,   , where the color of the curves represents the simulated device angle $\theta_\mathrm{sim}$ as introduced in (a). We notice that the magnetization aligns with the external field direction $\varphi_\mathrm{H,sim}=45^\circ+\theta_\mathrm{sim}$ at high field and rotates towards the nearby cubic anisotropy direction ($90^\circ$, $0^\circ$, and $-90^\circ$) as the external field is lowered. When the external field $\varphi_\mathrm{H,sim}$ is in the direction of cubic hard axis ($\theta_\mathrm{sim}=0^\circ, -90^\circ$), the magnetization direction $\varphi$ is locked in that direction (i.e., $\varphi=\varphi_\mathrm{H,sim}$) until the lowest frequency is reached, after which $\varphi$ rapidly changes to the nearest easy-axis direction ($\pm90^\circ$, not $0^\circ$ as uniaxial easy axis makes the former favorable than the later for $\theta_\mathrm{sim}=0^\circ$ or $-90^\circ$). (c) Magnetization in the simulated sample for $\theta_\mathrm{sim}=0^\circ$ for external field $H=260$~Oe. The magnetization direction $\varphi$ along with other related directions is illustrated at the bottom. 
    }
    \label{fig:fig4}
\end{figure}

 We analyze the STFMR results considering the free energy expansion and finding the resonance condition for our in-plane geometry as follows \cite{Aktas2007}:
\begin{equation}
H_0 sin(\varphi-\varphi_\mathrm{H})+\frac{1}{2} H_1 sin~ 4\varphi-H_\mathrm{u} sin~ 2\varphi=0
    \label{eq.  -a}
\end{equation}

    \begin{multline}
        \left(\frac{\omega_0}{\gamma}\right)^2=[H_0 cos(\varphi-\varphi_\mathrm{H}) +\frac{1}{2} H_1 (3+cos~ 4\varphi)\\+ 4\pi M_\mathrm{eff}-2H_\mathrm{u} cos^2\varphi  ] [ H_0 cos(\varphi-\varphi_\mathrm{H})\\+2H_1 cos~ 4\varphi-2H_\mathrm{u} cos~ 2\varphi] 
        \label{eq.  -b}
    \end{multline}
Here, $H_0$ is the external resonance field, $H_1$ and $H_\mathrm{u}$ are the cubic and uniaxial anisotropies, and $\varphi$ and $\varphi_\mathrm{H}$ are the direction of the magnetization and the direction of the external field, respectively [see Fig.~\ref{fig:fig1}]. The appearance of two resonance peaks can be understood by considering that the cubic hard direction is determined by an energy unstable equilibrium state superimposed with an energy gradient from the uniaxial anisotropy. When magnetic field is applied in that direction, it lowers the energy in that direction. As a result, the orientation of the magnetization is locked in the direction of external field [see Fig.~\ref{fig:fig4}(b)] and the resonance frequency drops with reduced magnetic field. When the magnitude of the energy associated with the external field is comparable to the energy gradient from the uniaxial anisotropy, the magnetization direction starts rotating to the easy direction. As the magnetization rotates towards the easy axis the effective field $H_\mathrm{eff}$ increases, making it possible to meet the resonance condition at two different external field values \cite{Papaioannou_APL2013} [see Fig.~\ref{fig:fig4}(a) and (b)]. 

The FMR condition is modeled based on the set of equations, Eqs.~(\ref{eq.  -a}) and (\ref{eq.  -b}). In the following, we describe the fitting procedure to experimental STFMR data we use to extract the magnetic parameters including  saturation magnetization $M_\mathrm{s}$ and anisotropy fields $H_\mathrm{u}$ and $H_1$. The equations have two independent variables, the external field $H_0$, and the external field angle $\varphi_\mathrm{H}$, while the static equilibrium magnetization direction $\varphi$ is a hidden variable that cannot be eliminated by analytically solving Eq.~(\ref{eq.  -a}). Therefore, we numerically solve Eq.~(\ref{eq.  -a}) for each pair of $H$ and $\varphi_\mathrm{H}$ to obtain the equilibrium orientation $\varphi$. This result is then used in Eq.~(\ref{eq.  -b}) to find the resonance frequency. An optimization process is implemented to find the global minimum of the least square residue \cite{newville_matthew_2014_11813} for the fitting parameters: saturation magnetization $M_\mathrm{s}=2226 \pm 5$~Oe, crystalline anisotropy fields $H_\mathrm{1}=215\pm 1$~Oe and $H_\mathrm{u}=4.4\pm 0.1$~Oe. The solid lines in Fig. \ref{fig:fig3} show the fits to the experimental data STFMR data (solid dots).

A comparison of our magnetic parameters and the literature is presented in Tab.~\ref{fit table}. The saturation magnetization we find is higher than the bulk value of Fe \cite{Goryunov_Phys.Rev.B1995}, while the anisotropy constants are slightly lower than the literature values \cite{Keller-prb-2017}.
Enhancement of magnetization in 3d/5d multilayers have been reported previously, e.g., \cite{Antel_1999,Simopoulos,CELINSKI,Wilhelm_PRL}. The effect is usually attributed to the narrowed d bands and localized electronic states. The latter emanates from the changes in the symmetry and the coordination number of ferromagnetic atoms located at or near a surface or a metal-metal interface.
Particularly in the Fe/Pt system, the interplanar distance and the Fe-Pt hybridization of the electronic wave functions are considered as the key factors for this enhancement \cite{Antel_1999,Simopoulos}.

\begin{table}[t]
\centering
\begin{tabular}{|c | c | c |} 
 \hline
 Parameter & Value (This work) (Oe) & Value (Literature) (Oe) \\ 
 \hline
 $H_s$ & 2226 $\pm$ 5 & 1700\cite{Goryunov_Phys.Rev.B1995} \\ 
 $H_1$ & 215 $\pm$ 1 & 260\cite{Keller-prb-2017}\\
 $H_u$ & 4.4 $\pm$ 0.1 & Negligible\cite{Keller-prb-2017}\\
 \hline
\end{tabular}
\caption{Magnetization parameters obtained by fitting experimental results to Eq.~(\ref{eq.  -b}) and comparison to literature values.}
\label{fit table}
\end{table}

We further verified our result through micromagnetic simulation using Mumax3 \cite{Vansteenkiste2014}. For the magnetization parameters we relied on the values obtained from the fits to the experimental data as summarized in Tab.~\ref{fit table}. Figure~\ref{fig:fig4}(a) shows the simulated resonance frequency vs. field plot, where an excellent agreement with the experimental FMR and STFMR results is found. 
Moreover, the micromagnetic simulations enable us to determine and visualize the magnetization direction with respect to the hard axis of the uniaxial anisotropy (as per our coordinate system) as a function of the external field for each simulated field angle, $\theta_\mathrm{sim}$. As is apparent from Fig.~\ref{fig:fig4}(b) the magnetization tries to align with the external field at a high external field 
($\varphi_\mathrm{H,sim}=45^\circ+\theta_\mathrm{sim}$) and rotates continuously towards the nearby easy anisotropy direction ($90^\circ$, $0^\circ$, or $-90^\circ$) as the external field reduces (note that the curves are slightly ``bent'' as the field is lowered). By comparing Fig.~\ref{fig:fig4}(a) and (b), we see that the resonance frequency increases as the orientation of the magnetization moves away from the direction of the external field.  However, when the external field is along the cubic hard direction, for $\theta_\mathrm{sim}=0^\circ$ or $\theta_\mathrm{sim}=-90^\circ$, the magnetization stays locked in the direction of the external field ($\varphi_\mathrm{H,sim}=45^\circ$ and $-45^\circ$, respectively) while the resonance frequency steadily reduces. 
After it reaches the lowest resonance frequency at about ~$450$ Oe [see Fig. \ref{fig:fig4}(a)], the magnetization attempts to quickly align in the direction of the easy axis ($\varphi_\mathrm{H,sim}=45^\circ$ and $-45^\circ$, respectively). This magnetization re-alignment away from the external field direction is accompanied by a fast increase in frequency [see Fig. \ref{fig:fig4}(a)]. 

In summary, we demonstrated the generation and detection of STFMR in micropatterned epitaxial Fe/Pt bilayers grown by molecular beam epitaxy. Using an analytical formalism we extract the material parameters including saturation magnetization, uniaxial, and cubic anisotropies. It is found that saturation magnetization is larger than the literature value for bare Fe thin films, which is likely due to  induced magnetic moments mediated by the presence of the Pt capping layer and the microstructuring of the devices. Micromagnetic modeling using Mumax3 revealed that the magnetization rotates in the direction of the nearby cubic anisotropy direction as the field is lowered to minimize the total energy. Thus, the condition for maximum STFMR is no longer fulfilled and the signal intensity decreases in that field range. Our results show that STFMR can reveal these magnetic anisotropies in individual microstructure devices -- a challenging task for conventional $rf$ spectroscopy techniques such as broadband FMR due to their relatively signal strength for microstructures.

\begin{acknowledgments}
We thank Dr. Hang Chen and Dr. John Xiao for assistance with the broadband FMR measurements. This work was supported by NSF through the University of Delaware Materials Research Science and Engineering Center DMR-2011824.
\end{acknowledgments}

The data that support the findings of this study are available upon reasonable request.

\bibliography{biblio}

\end{document}